\documentclass[preprint,12pt]{elsarticle}
\usepackage{graphicx} 
\usepackage{subfigure} 
\usepackage{epsfig,amssymb,a4}

\newcommand{\beq}{\begin{equation}}
\newcommand{\eeq}{\end{equation}}
\newcommand{\beqa}{\begin{eqnarray}}
\newcommand{\eeqa}{\end{eqnarray}}

\newcommand{\ben}{\begin{enumerate}}
\newcommand{\een}{\end{enumerate}}
\newcommand{\bit}{\begin{itemize}}
\newcommand{\eit}{\end{itemize}}
\newcommand{\bce}{\begin{center}}
\newcommand{\ece}{\end{center}}

\graphicspath{{./figures/}}

\journal{arXiv}
\begin{document}

\begin{frontmatter}
 
\title{PIV in stratified gas-liquid flow in a horizontal pipe using water droplets as tracers in the gas-phase}

\author[Ay]{A.A. Ayati}

\address[Ay]{Department of Mathematics, University of Oslo, N-0316 Oslo, Norway}
\ead{awalaa@math.uio.no}

\author[Ko]{J. Kolaas}
\author[Je]{A. Jensen}
\author[Jo]{G. Johnson}

\begin{abstract}

Simultaneous Particle Image Velocimetry (PIV) measurements of stratified turbulent air/water flow in a horizontal pipe have been performed using water droplets as tracers in the gas-phase. The use of water droplets as tracers ensures that the water surface tension remains unaffected and thus allows small scale interfacial structures, such as capillary waves to occur naturally. Experiments have been conducted in a 31 m long, 100 mm diameter PVC pipe using air (density $\simeq$ 1.20 kg/m$^3$ and viscosity 18.4 $\mu$Pa$\cdot$s) and water (density 996 kg/m$^3$ and viscosity 1.0 mPa$\cdot$s) as test fluids. For the purpose of validation of the experimental set-up and the suggested seeding technique, single-phase measurements of both air and water were compared to each other and to DNS results provided by "Wu X. and Moin P., 2008, A direct numerical simulation study on the mean velocity characteristics in turbulent pipe flow, J. Fluid Mechanics, Vol. 608.", showing very good agreement. The two-phase measurements are presented in terms of mean- and rms-profiles. These measurements offer a qualitative demonstration of the behavior of the interfacial turbulence and its correlation with the various interfacial flow patterns. 

The observations made in this paper are in agreement with the conclusions drawn from the DNS study of "Lakehal D., Fulgosi M., Banerjee S. and De Angelis, Direct numerical simulation of turbulence in a sheared air water flow with a deformable interface, 2003, J. Fluid Mechanics, Vol. 482.". The present results may eventually provide a better explanation to many important phenomena related to the physical characteristics of stratified two-phase flow such as scalar mixing between phases, and to challenges related to its modeling.

\end{abstract}


\end{frontmatter}

\section{Introduction}
Stratified two-phase flow is a flow regime that occurs when the velocity of each of the phases is relatively low. In such flows the inertia forces are not large enough to generate large waves which may lead to instabilities that usually initiate the onset of intermittent flow regimes like slug flow or dispersed flow. Nevertheless, stratified flow is frequently encountered in petroleum transportation pipes, in the nuclear and process industries under steady or transient conditions. In the natural gas industry, gas/liquid flow is the dominating two-phase combination and is mainly present as a gas/condensate or gas/water mixture. The liquid condensation of natural gas is an inevitable process that occurs due to the temperature and pressure changes that exist along the pipes. In off-shore gas fields, the raw production is often transported in multiphase pipelines before it reaches a processing unit. These lines lie at the bottom of the sea in horizontal and near-horizontal positions. Hence, a better understanding of the flow characteristics of gas/liquid flow in horizontal pipes is needed for proper design and operation of pipelines, see e.g Mokhatab (2006).

The key design parameters are the pressure drop and average in situ holdup and velocities. Their prediction has traditionally been based on greatly simplified representation of the flow structure where both phases are treated as one-dimensional bulk flows (two-fluid 1D model), see Ullmann and Brauner (2006), Schulkes (2010) and Johnson (2005). However, the application of the two-fluid model relies on the availability of interfacial information such as the interface shape and closure relations for the wall shear and interfacial shear stresses. These closure relations should cover both the systems parameters (e.g. fluids flow rates, physical properties and pipe characteristics) and flow related parameters in either phases. In the most common approach, an empirical correlation for the interfacial friction factor or the interfacial friction term is obtained from experimental data. Among others, the models proposed by Andreussi and Persen (1986), Andritsos and Hanratty (1987), and Biberg (2007) are based on this method. It should be worth mentioning that the latter model is incorporated in the latest version of the flow assurance simulator, OLGA. 

Although flow characteristics such as the pressure drop and liquid hold-up have been extensively studied, our understanding of the turbulent flow structure in stratified two-phase flows is still very limited. The local structure of turbulence is an important factor in the transport and mixing of mass, momentum and energy between the phases. Therefore, a better understanding of the relation between different flow patterns and their belonging turbulence profiles might be the key to more accurate mathematical models. However, the details of the turbulence structure near the gas-liquid interfaceare difficult to obtain using conventional instrumentation such as hot-wire or hot-film anemometery, as the probes can seriously interfere with the water surface. Fabre et al. (1987) have presented one of few successful LDA measurements of turbulence parameters close to the interface of wavy stratified flow.

The present work is an experimental study on air/water stratified flow in a horizontal pipe aiming to eventually provide accurate interfacial turbulence data to Biberg's mathematical model (2007). Particle Image Velocimetry (PIV) technique has been utilized for the simultaneous measurements of both phases, whereas traditionally, only liquid phase measurements have been documented, see Carpintero-Rogero et al. (2006). PIV is a non-intrusive measuring technique that provides instantaneous two-dimensional velocity and turbulence fields by tracing the flow with passive particles and acquiring images of these particles with the help of a high resolution camera, a high power double-pulsed laser and a signal processing system based on advanced cross-correlation methods, see Sveen and Cohen (2004) and Westerweel (2008). 

For most PIV experiments it is desirable that seeding particles are non-toxic, non-corrosive, non-abrasive, non-volatile and chemically inert. Traditionally, oil droplets or solid particles have been used as tracers in gas single-phase flows. Melling (1997) presents a clear and concise overview on existing PIV studies categorized by the seeding particles 
that have been used. He concludes that seeding with liquid droplets offers the advantage of a steadier production rate than is normally feasible with solid particles. But once the studied flow contains an additional phase, which in this case is water, seeding with oil droplets can interfere chemically with the water surface tension. Small scale interfacial 
structures such as capillary waves are dependent on the water surface tension and these structures are the first steps towards more dramatic flow regimes. The novelty of this work lies in the use of water droplets of $1\sim 5 \mu m$ in diameter as tracers in the gas phase and the acquirement of velocity profiles from both phases simultaneously, providing better understanding of the effects of interfacial turbulence.

Several tests have been carried out in order to legitimize the gas phase seeding technique. Single-phase measurements of the gas flow were conducted at $Re = 44000$ and compared to DNS results obtained by Wu and Moin (2008), and at lower $Re$-numbers and compared to water single-phase measurements. Both the single- and two-phase results are presented in this paper. 

The paper is structured as follows: the experimental set-up and measurement technique which includes the single phase measurements are described in section \ref{sec::Experimental set-up}. Results from two-phase flow measurements are presented in terms of velocity and turbulence profiles in section \ref{sec::Two-Phase}. Furthermore, in section \ref{sec::STA} an analysis based on the gas-phase turbulence is presented showing that there exists a clear correlation between each interfacial flow pattern (stratified sub-regime) and its corresponding axial turbulence structure. Finally, a summary containing conclusive remarks is found in section \ref{sec::conclusion}.

\section{Experimental set-up and measuring technique}
\label{sec::Experimental set-up}

\begin{figure}[htp]
\centering
\includegraphics[angle=0,scale=1]{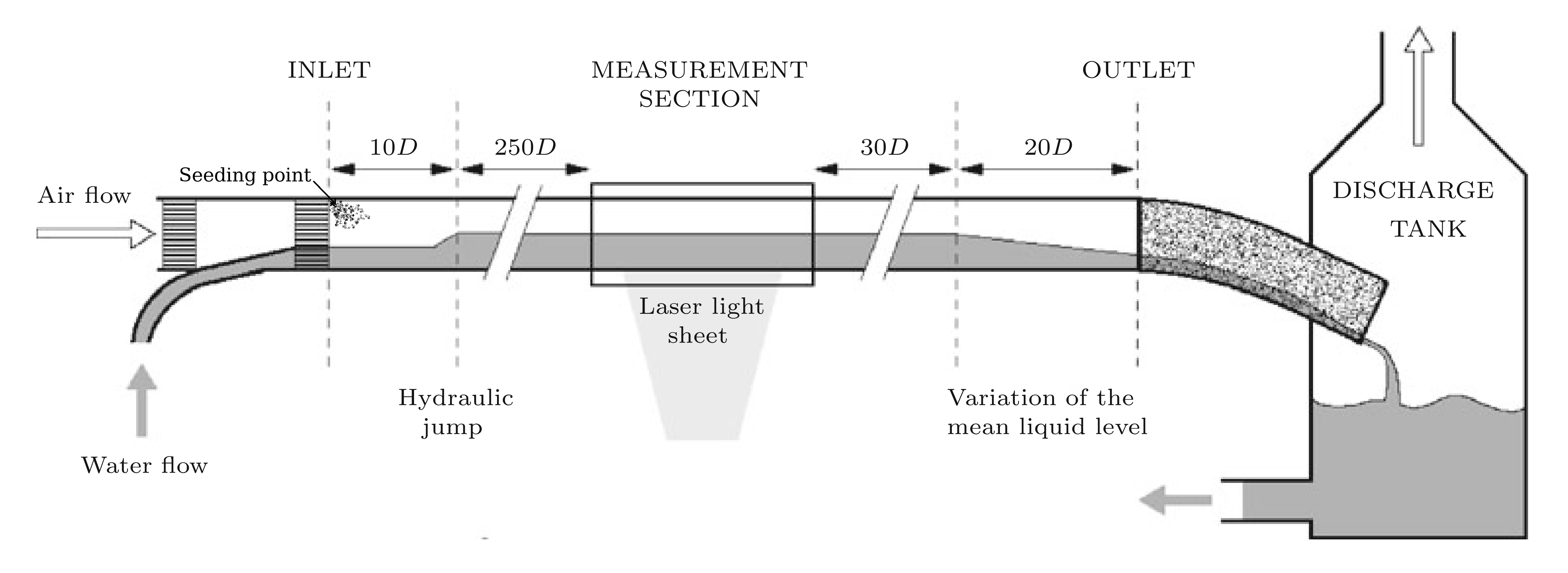}
\caption{Schematic view of the pipe loop used in the present work.}
\label{Fig::PipeLoop}
\end{figure}

The PIV experiments were conducted in a horizontal 31 m PVC pipe with an internal diameter D = 10 cm. The pipe consisted of adjacent sections, each with a length of 3.5 m connected by annular joints that ensured tightness. All joints were rigidly attached by collars to vertical beams that supported the whole structure. The test fluids were air and water at atmospheric pressure with an average temperature of 22$^\circ$C. Figure \ref{Fig::PipeLoop} shows the disposition of the pipe elements. 

Water was injected at the pipe bottom through a 5 cm I.D. tee branch. Honeycomb flow straighteners were placed right before and after the contact point between the liquid and gas phases. At the outlet, the pipe discharged into a separating tank at atmospheric pressure in which both the water and air were recirculated from the bottom and top exits of the tank, respectively. 

Furthermore, water was circulated with a maximal volumetric flow rate of 90 m$^3$/h, and a frequency-regulated fan produced the airflow. The water and air mass flow rates were measured with Coriolis flow meters. Bulk velocities were calculated using a density of 997 kg/m$^3$ for water and around 1.2 kg/m$^3$ for air, corrected with the temperature at the measurement point. Reynolds numbers in the gas and water single-phase based on the pipe diameter could reach up to 48$\cdot$10$^3$ and 25$\cdot$10$^3$, respectively. For more details about the facilities, see Sanchis and Jensen (2011).

PIV velocity measurements in a vertical plane were performed in a test section located 260D downstream from the pipe inlet and 50D upstream from the outlet. A high power double-pulsed Nd:YAG laser of 135 mJ provided the illumination, and images were recorded with two PCO.4000 cameras (only one for the single-phase measurements) at a rate of one frame-pair per second. The cameras were placed 20 cm under and above the pipe center line at an upward and downward angle of 20$^\circ$ in order to map both phases entirely including the interface. The water phase was seeded with polyamid particles of 50 $\mu m$. The gas phase was seeded with water droplets of various sizes which were injected at the inlet of the pipe loop. A "natural selection" process ensured that only the droplets with the right sizes (mainly 1 to 5 $\mu m$) remained in flow reaching the test-section after 21m (210D). Heavy droplets quickly descended and formed a thin liquid water stream that flowed alongside the pipe bottom. This mechanism caused some measuring deviations in gas single-phase experiments, but it did not have any effect on the two-phase experiments,as the water stream just blended with the water-phase. The deviations caused in the single phase flows are discussed later.   

\begin{figure}[ht!]
\centering
\includegraphics[width=\columnwidth, height=0.5\textheight]{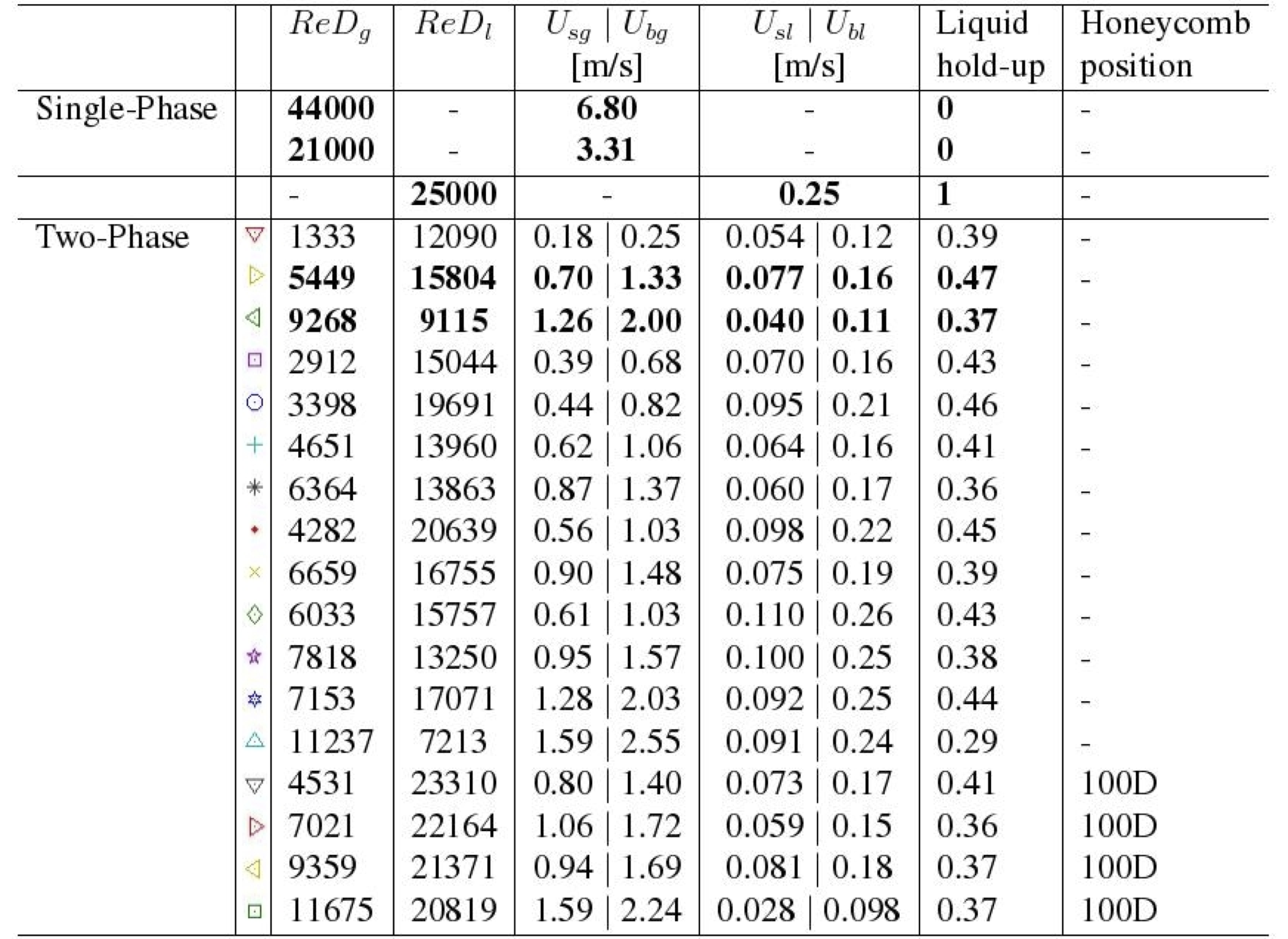}
\caption{\small An overview of all experimental cases that are discussed in this paper. Each experiment is characterized by its hydraulic diameter based $Re$-number, superficial and bulk velocity, 
average liquid hold up and the position of a honey comb flow straightener. The bold numbers represent the cases that are presented by their velocity profiles 
in sections \ref{sec::Single-Phase} and \ref{sec::Two-Phase}.} \label{Tab::Re-numbers}
\end{figure}

In this paper, a set of 3 single-phase flows and 17 stratified air/water two-phase flows of varying flow rates are presented. All experiments, characterized by their single or mixture flow rates, are classified in figure \ref{Tab::Re-numbers}. In each case, between 213 and 568 image-pairs, 1500 $\times$2600 pixels in resolution, were recorded by each camera. Sub-window sizes varied between 100$\times$32 and 64$\times$32 depending on the flow velocity and time separation.

The presented experiments were conducted over a period of three months in which the calibration procedures concerning amongst others, the pressure probe and pixel to physical coordinates transformation were repeated several times. 

In the two-phase experiments, relatively low mixture velocities were combined in such a manner that the interface pattern varied from smooth to small amplitude 2D waves. Distinction between the different interface patterns was based upon visual observation. Espedal (1998) and Fernandino and Ytrehus (2006) provide both visual and quantified interface descriptions for stratifiedair-water pipe and channel flows at different flow rates. These descriptions were found to agree very well to the observations made in this study. This is discussed further in section \ref{sec::STA}. Some of the two-phase cases were conducted with an additional honey comb flow straightener inside the main pipe section upstream of the test section. This was done because the honeycomb acted as wave-damper and offered the system a broader interval of flow rates that generated non-wavy flows. Several tests were conducted in order to make sure that distance between the honeycomb and the test section was large enough such that the flow disturbance caused by the honeycomb was dissipated. Eventually, the minimum distance of $10 m$ (100D) upstream of the test section was chosen in four cases.

\begin{figure}[ht!]
\centering
\includegraphics[width=0.8\columnwidth, height=0.4\textheight]{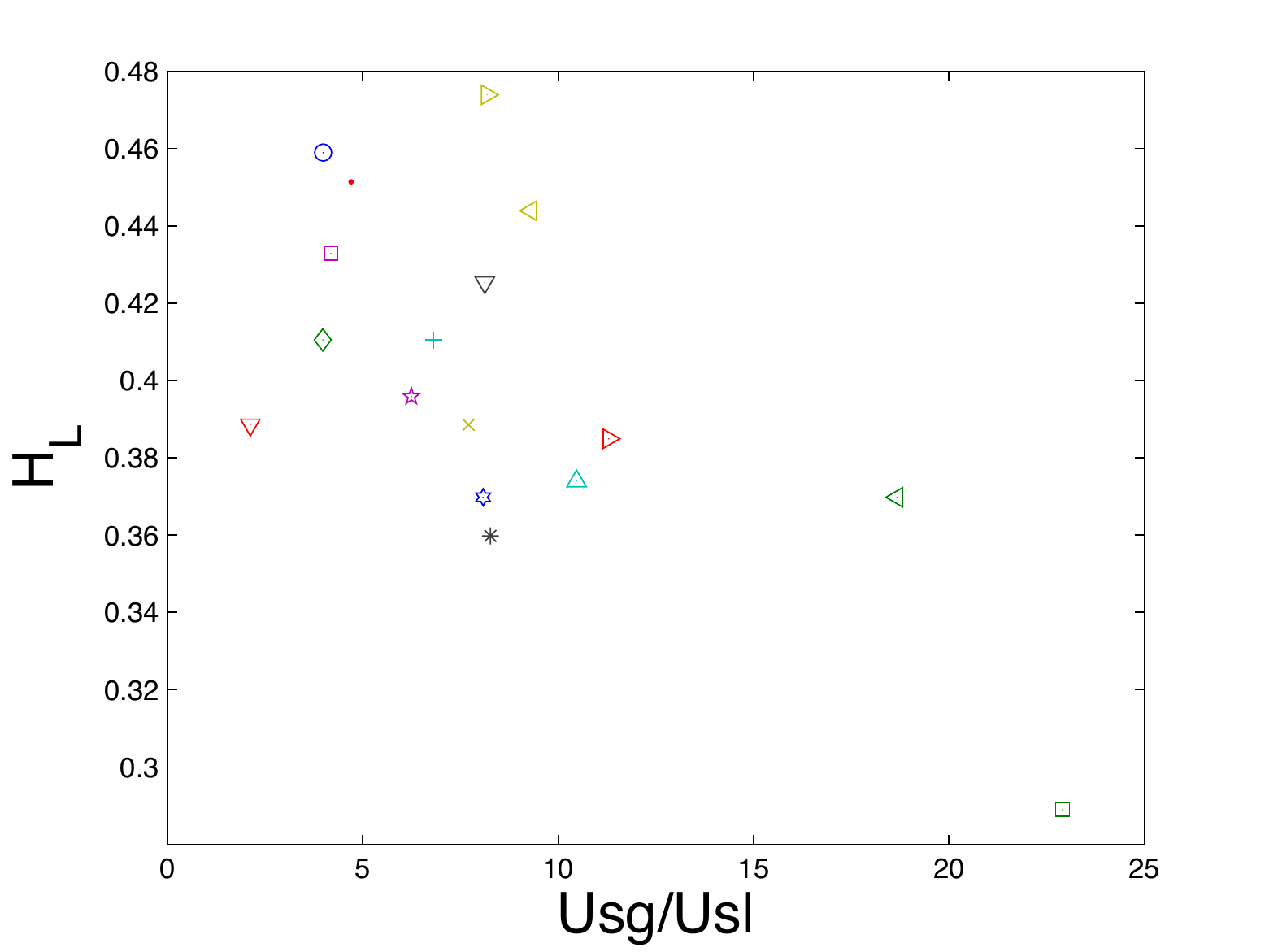}
\caption{\small The average liquid hold-up $H_L$ plotted against $U_{sg}/U_{sl}$.} \label{Fig::TurbAnalysis_ReDg_vs_HL}
\end{figure}

Figure \ref{Fig::TurbAnalysis_ReDg_vs_HL} shows the average liquid-hold up of each experiment plotted against the superficial velocity fraction $\frac{U_{sg}}{U_{sl}}$.
Most of the experiments were conducted with a liquid hold-up between $0.35$ and $0.5$ except from the wavy case represented by the green square. Since the flow straightener allowed for higher gas flow rates to produce non-wavy flows, the liquid height was forced to a lower position leading to higher bulk velocities in the liquid phase. 

\subsection{Interface Detection and Validation}
\begin{figure}[ht!]
\centering
\includegraphics[width=\columnwidth, height=0.8\textheight, angle=90]{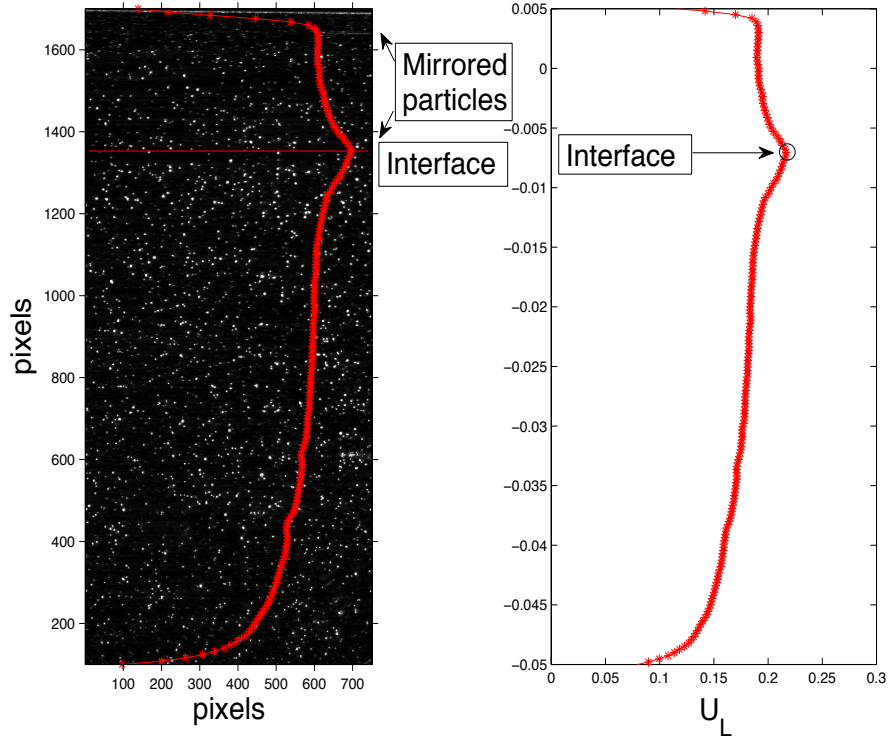}
\caption{\small Interface detection method.} \label{Fig::InterfaceDetection}
\end{figure}

In the case of two-phase flow in pipes, the liquid surface is curved as a consequence of the roundness of the pipe. Thus, the vertical position of the interface at the centerline is actually lower than at the pipe wall. This detail forced the rather complicated measuring set-up (consisting of two cameras) described in the previous section. Figure \ref{Fig::InterfaceDetection} shows a typical liquid phase PIV image acquired from the lower camera. Due to the inclination of the camera, a part of the upper region of the liquid phase is reflected onto the water surface. Thus, mirrored tracer particles appear between the real center line interface and the artificial interface, which in fact is just the area of intersection between the water surface and the pipe wall. See Arnaud and Jensen (2011) for more details.

This particular feature, typical of two-phase pipe flow, initially seems as an additional practical challenge, but this is not the case. Once processed, the liquid phase images provide a mean axial velocity profile that consists of a characteristic bump as seen from figure \ref{Fig::InterfaceDetection} (right). The peak of this bump gives the time-averaged vertical position of the real centerline interface with quite good accuracy, especially in cases with smooth interface and waves with small amplitudes. Furthermore, the interface position or the liquid height, is used to calculate the hydraulic diameter $D_{H,f}$, the hydraulic Reynolds number $ReD_f$ and the local bulk velocity $U_{bf}$ of each phase:

\beq
U_{b,f} = \frac{\dot{m}_f}{A_f \rho_f} = U_{sf}\frac{A}{A_f}.
\eeq  

The hydraulic Reynolds number is defined as

\beq
\label{Eq::ReD_f}
ReD_f  = \frac{\rho_f U_{b,f} D_{H,f}}{\mu_f}.
\eeq

in which $D_{H,f}$ is the hydraulic diameter

\beq
D_{H,l} = \frac{4A_l}{s_l}, \hspace{2 cm} D_{H,g} = \frac{4A_g}{s_g + s_i}.
\eeq

\begin{figure}[ht!]
\begin{center} 
\begin{minipage}[t]{0.48\columnwidth}
\centering
\includegraphics[width=\columnwidth, height=0.3\textheight]{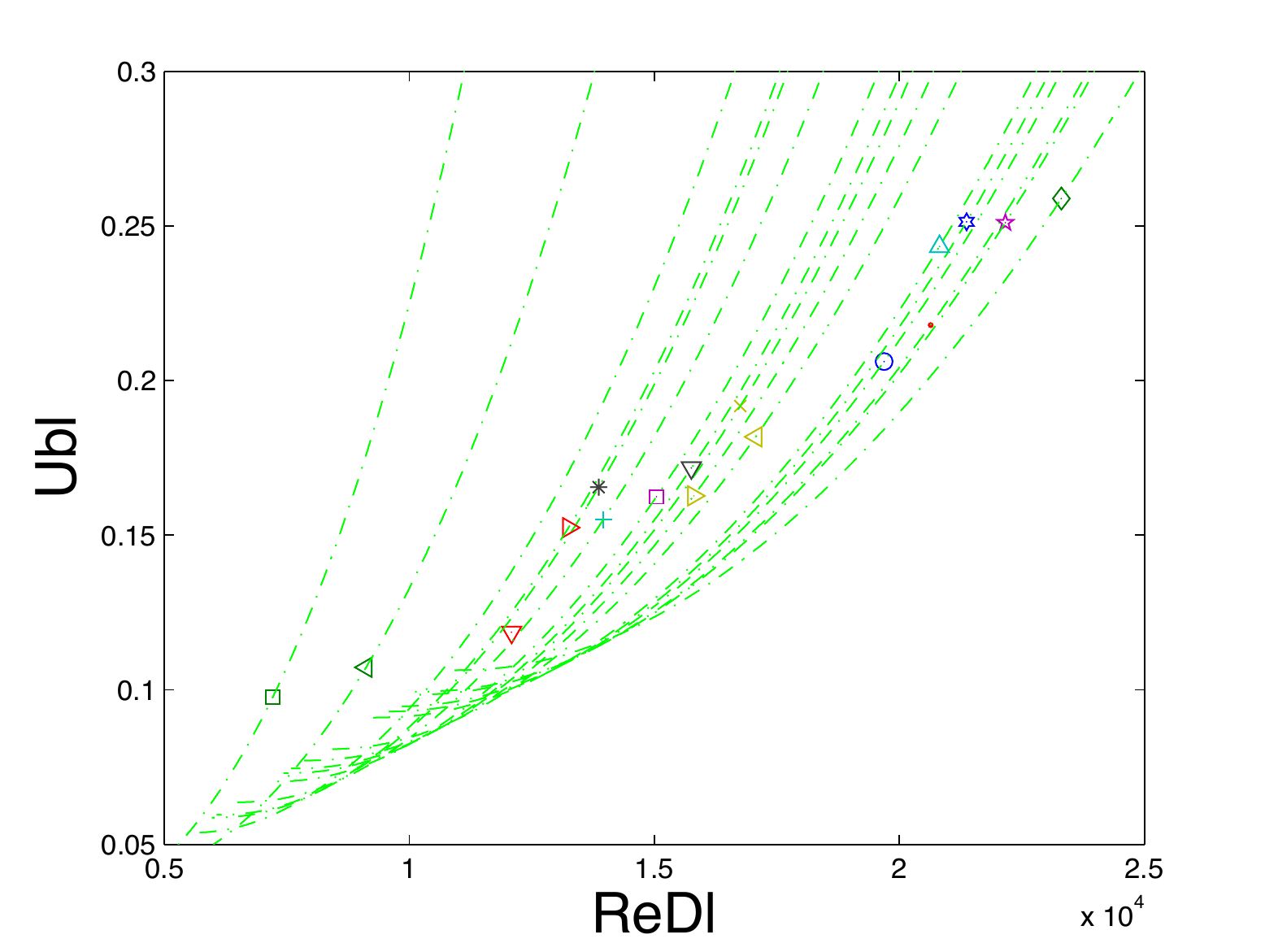}
\caption{\small The water mean velocity $Ub_L$ plotted against $ReD_l$. The green lines show the theoretical relation between them for varying $H_L$ and given $U_{sl}$.} \label{Fig::TurbAnalysis_ReDl_vs_Ubl}
\end{minipage}
\hspace{0.15cm}
\begin{minipage}[t]{0.48\columnwidth}
\centering
\includegraphics[width=\columnwidth, height=0.3\textheight]{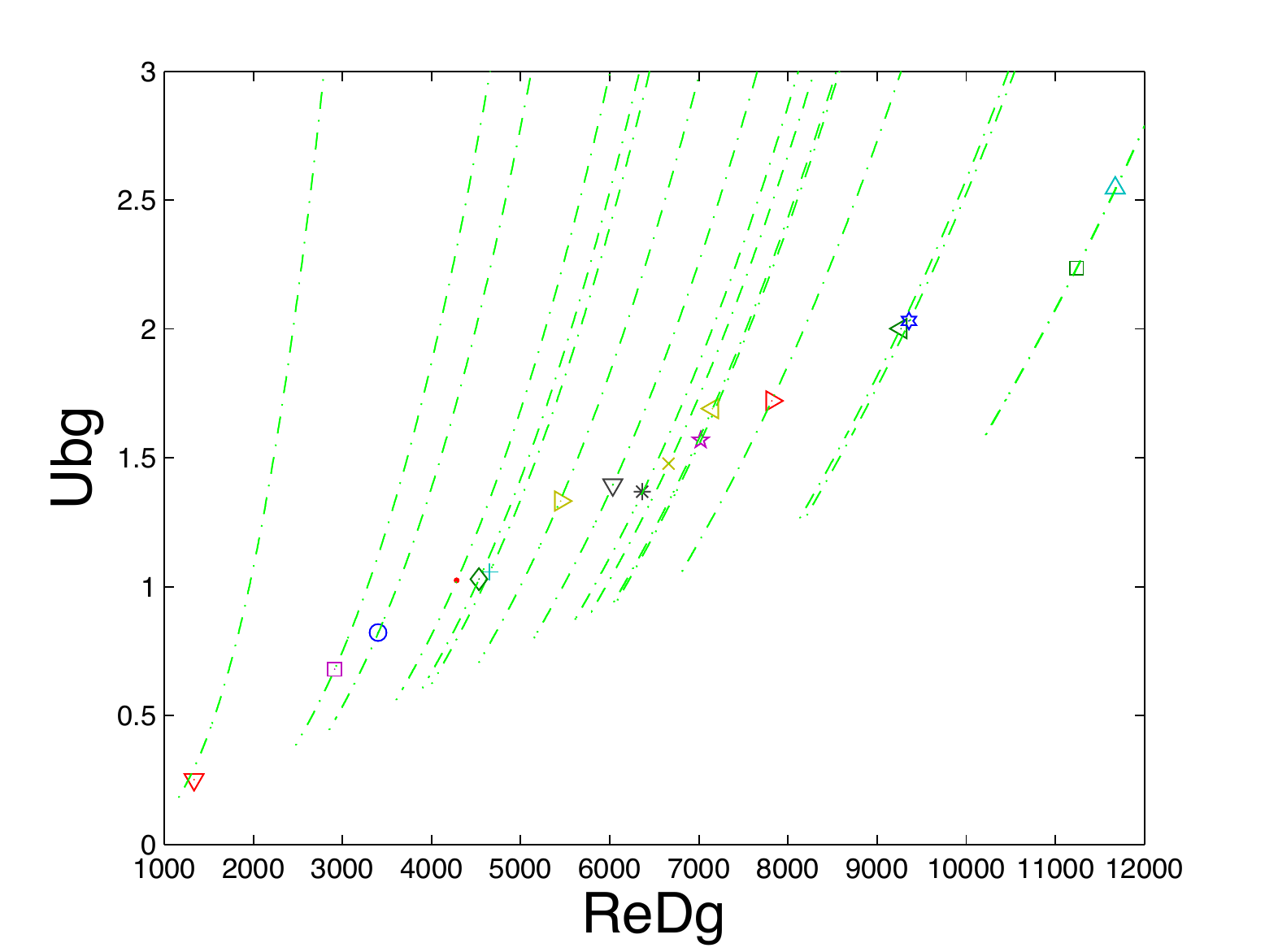}
\caption{\small The gas mean velocity $Ub_g$ plotted against $ReD_g$. The green lines show the theoretical relation between them for varying $H_L$ and given $U_{sg}$.} \label{Fig::TurbAnalysis_ReDg_vs_Ubg}
\end{minipage}
\end{center}
\end{figure}

Figures \ref{Fig::TurbAnalysis_ReDl_vs_Ubl} and \ref{Fig::TurbAnalysis_ReDg_vs_Ubg} show the relationship between the hydraulic $Re$-numbers of each phase $ReD_f$ and their corresponding local mean velocities $U_{b,f}$ for all two-phase flow experiments presented in this paper. The green lines express the theoretical relation between $ReD_f$ and $U_{b,f}$ defined in equation \ref{Eq::ReD_f} for a given $U_{sl}$ and varying liquid heights. All the theoretical lines intersect with their corresponding experimental cases. Thus, this can be regarded as a validation of the flow meters and the interface detection method. 

\subsection{Single-phase flow measurements}
\label{sec::Single-Phase}

Reliability of the current experimental set-up and the suggested gas flow seeding technique may be demonstrated partially through comparison with existing numerical or experimental data. For this purpose, PIV measurements of fully developed single phase flows were performed with both air and water at three different $Re$-numbers: $Re_g = 44000, 21500$ and $Re_l = 25100$. The results are compared to the DNS data provided by Wu and Moin (2008). They carried out a DNS simulation of fully developed incompressible turbulent pipe flow at $Re= 44 000$ with second-order finite-difference methods on 630 million grid points. The lower $Re_g$ case is compared to the liquid single phase flow. The measurements are presented in terms of normalized mean horizontal velocity profile $\frac{\overline U}{U_b}$, horizontal and vertical rms-profiles, 
$\frac{u'}{u^*}$ and $\frac{v'}{u^*}$, and $Re$-stress profiles $\frac{u'v'}{u^{*2}}$. Where $U_b$ is the bulk velocity calculated from the measured mass flow rate $\dot{m}$ and $u^*$ is the friction velocity calculated from the measured pressure drop $\Delta P$ using a momentum balance approach. Respectively:

\beq
\label{Ub}
U_b = \frac{\dot{m}}{\rho A}.
\eeq 
and

\beq
u^* = \sqrt{\frac{\tau_w}{\rho}}, \hspace{5mm} \tau_w = \frac{\Delta P D}{4L}.
\eeq

\begin{figure}[ht!]
\begin{center}
\includegraphics[width=\columnwidth, height=0.5\textheight]{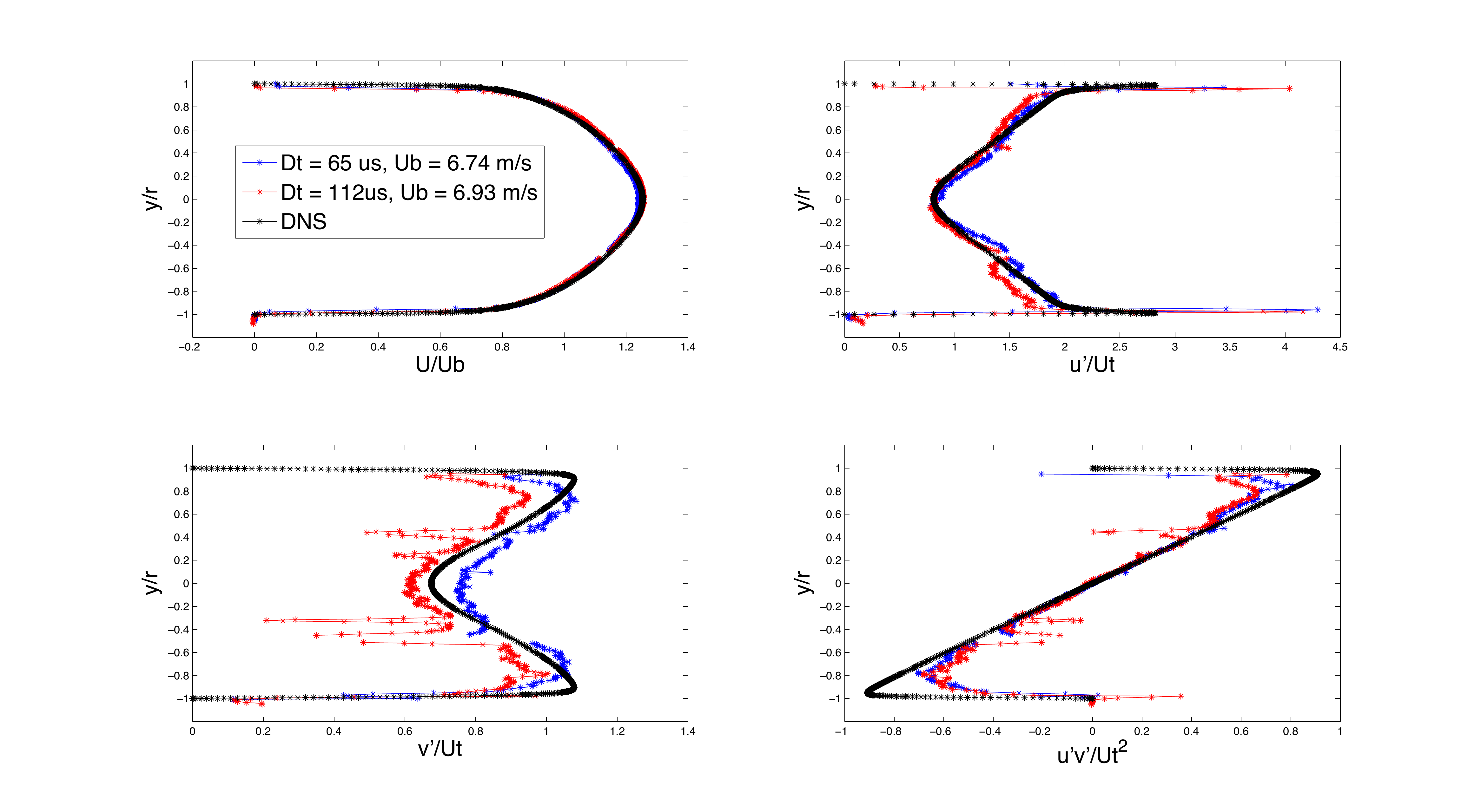}
\caption{\small Normalized $\bar U$ (top left), $u'$ (top right), $v'$ (bottom left) and $u'v'$-profiles of two gas single phase PIV experiments at $Re_g = 44000$ with 
$\Delta t = 65\mu s$ and $112 \mu s$ are compared to DNS data from Wu and Moin (2008).} \label{Fig::SinglePhaseGas_Re44k_Dt65_112_U_u_v_uv}
\end{center}
\end{figure}

Figure \ref{Fig::SinglePhaseGas_Re44k_Dt65_112_U_u_v_uv} shows a set of two gas single-phase '$Re=44k$'-experiments in which the time separation $\Delta t$ between each PIV image-pair was set to $\Delta t = 65 \mu s$ and $\Delta t = 112 \mu s$. The corresponding maximum particle displacements in pixels were [$\Delta x = 17, \Delta y = 0.2$] for case 1 and [$\Delta x = 40, \Delta y = 0.35$] for case 2, respectively.

Not surprisingly, the stream wise pixel displacements are substantially greater than the wall normal displacements, which are, in fact, at a subpixel level. This signals that a lower degree of accuracy is expected in second order statistics of the vertical component. This is seen in the '$v'$'-profiles throughout this paper, which all look a somewhat noisy. Despite this, a relatively good qualitative agreement is seen from the $v'$ plot in Figure \ref{Fig::SinglePhaseGas_Re44k_Dt65_112_U_u_v_uv}. Also, it looks like the $v'$-profile is more sensitive to changes in the image-pair time separation, whereas the $\Delta t =65\mu$-case leads to higher overall wall normal turbulence level than the $\Delta t = 112 \mu s$. This gives us valuable information concerning the optimization of $\Delta t$, where the $Re$-number and the size of the field of view are parameters, amongst many others, that need to be carefully taken into account. 

\begin{figure}[ht!]
\begin{center}
\includegraphics[width=\columnwidth, height=0.5\textheight]{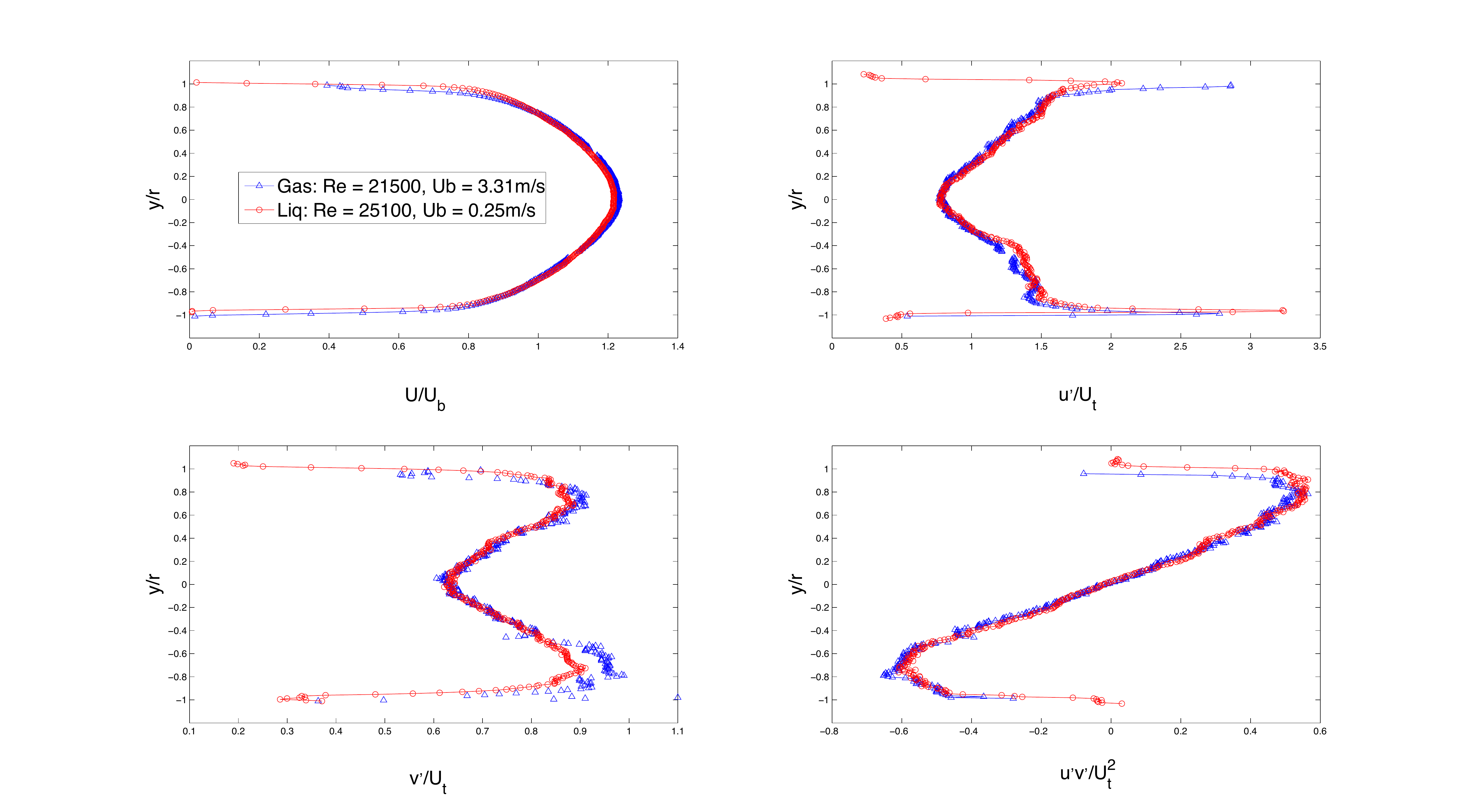}
\caption{\small  Normalized $\bar U$ (top left), $u'$ (top right), $v'$ (bottom left) and $u'v'$-profiles of gas and water single phase measurements at $Re_g= 21500$ (blue) and 
$Re_l = 25100$ (red).} \label{Fig::PIV_SinglePhase_G21k_L25k_vs_DNS}
\end{center}
\end{figure}

First order statistics of the gas single-phase, here $\bar U_g$, compare perfectly with both the DNS simulations and the $Re_l=21500$ liquid single phase experiment  
shown in Figure \ref{Fig::PIV_SinglePhase_G21k_L25k_vs_DNS}. The axial turbulence profiles also show satisfactory agreement in the main pipe region in both comparisons. However, close to the pipe walls, the PIV peaks exceed the DNS peaks by roughly $50\%$. This can be due to either measurement uncertainties related to levels of distortion 
near the pipe walls, or an under prediction in the DNS calculations. In the case of the first cause, the uncertainties should be of equal magnitude on both ends of the pipe.
Thus, the ratio between these peaks should reflect the nature of the flow, meaning that for a fully turbulent and symmetric flow, this ratio should be equal to 1. This is seen only in the $\Delta t=112 \mu s$-case and $Re_g=21500$. The asymmetry seen in the $\Delta t=65 \mu s$-case was actually caused by the water stream formed by the heavy tracer droplets, as explained earlier. The higher peak at the bottom is in accordance with the fact the water stream was wavy, and thus increased the roughness locally. This is actually a two-phase flow phenomenon and will be discussed more in detail later in this paper. The water stream was partially avoided in the other gas single-phase case. 

The asymmetry seen in the water case is reversed (top peak is lower than bottom peak) and was caused by a totally different situation. Occasionally, elongated air bubbles occupied a very thin layer at the top wall and changed the boundary conditions locally, leading to free surface flow. Hence, the averaged axial turbulence level was lower than in a pure single-phase flow.

It can be concluded from the presented single phase tests, that the PIV set-up with its gas seeding technique provides very good measurements in both phases. The few deviations that have been observed are due to explainable circumstances which, luckily, have no significance in two-phase flow experiments.  

\section{Two-Phase measurements}
\label{sec::Two-Phase}

\begin{figure}[ht!]
\centering
\includegraphics[width=\columnwidth, height=0.5\textheight]{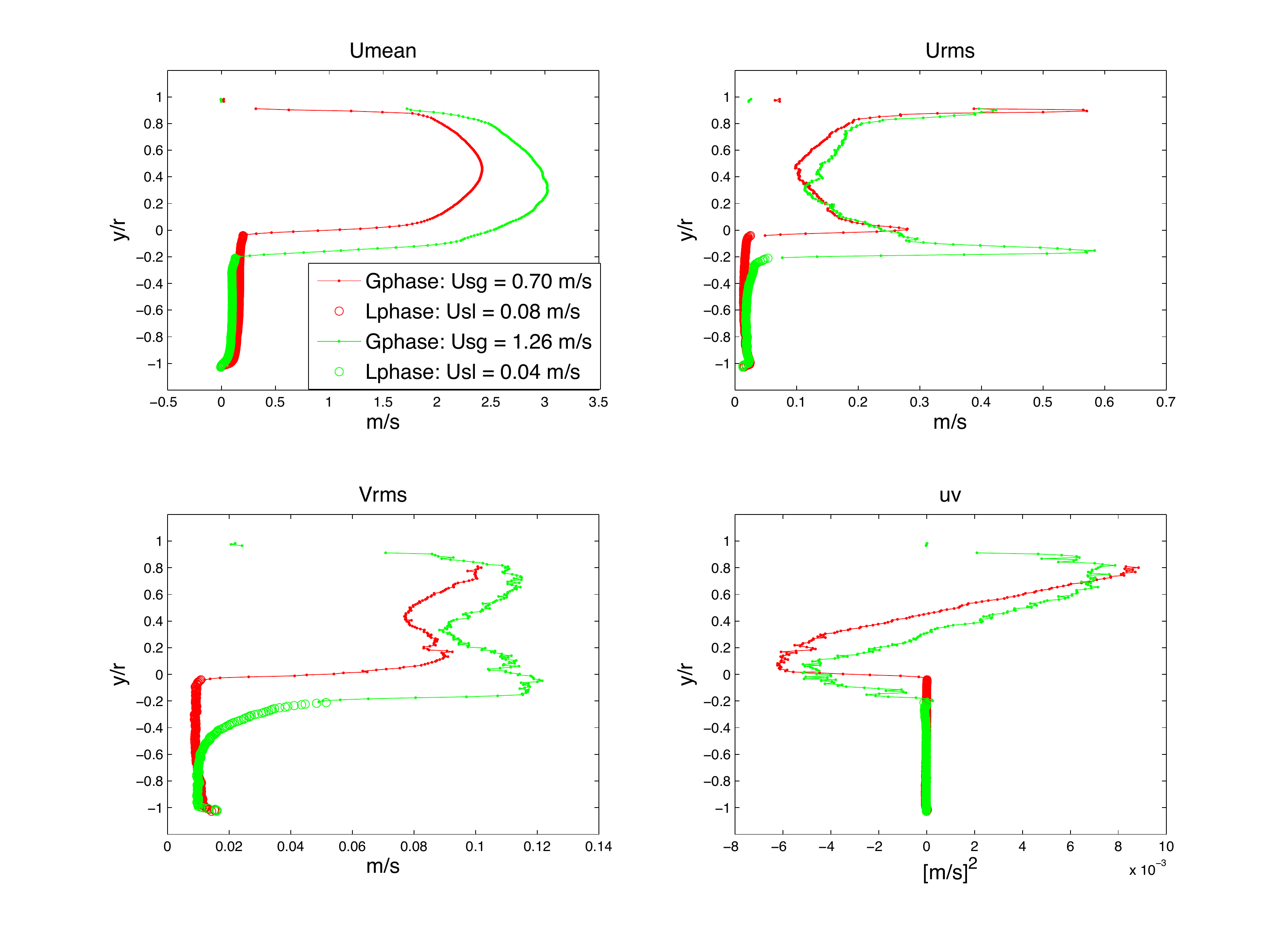}
\caption{\small Mean axial profile $\overline U$ (upper left), axial (upper right) and vertical (lower left) fluctuating velocity profiles $u'$ and $v'$ and finally $Re$-stress profile $u'v'$ 
for a flat smooth interface (red) and wavy interface (green). The thick lines indicate measurements in the water-phase.}
\label{Fig::PIV_2Phase_2Runs_Profiles_May2012}
\end{figure}

In this section a set of two stratified air/water flows are presented in terms of the vertical line mean and rms velocity profiles, in which the interface detection method described earlier has been applied to obtain smooth profile transitions between the phases, see Figure \ref{Fig::PIV_2Phase_2Runs_Profiles_May2012}. 
The physical parameters that characterize these experiments are listed in the table with bold characters, see figure \ref{Tab::Re-numbers}. The first case (red line) represents a stratified smooth flow, while the second case (green line) represents a wavy flow. The waves were typical long 2D gravity waves with relatively low amplitudes. 
 
The $\bar U$-profiles (top left plot) offer a global image of the flow characteristics of both cases. They show that the gas flow rate was increased considerably while the liquid rate was lowered in order to obtain the desired wavy regime. However, the most interesting effects of this sub-regime change are definitely seen in the turbulence profiles. In fact, a great deal of physics may be revealed just by studying the near-wall and near-interface peaks, especially in the axial turbulence profiles. 

In the smooth case, it is clear that the near-wall peak of the axial fluctuating gas velocity $u'_{g,w_p}$ is larger than the near-interface $u'_{g,i_p}$ peak. This feature is not surprising and can be explained quite easily through the boundary conditions that are imposed on the gas phase. Namely, no-slip condition both at a non-moving wall on one side, and a moving smooth interface on the other side. This situation is analogous to a turbulent couette flow in a semi-pipe with a moving bottom wall. The velocity gradient is much higher near the wall than near the interface leading to a higher turbulence level. 

Moreover, in the second case, where the flow is characterized by its wavy interface, a very different situation occurs. This time, the near-interface peak $u'_{g,i_p}$ is larger than both its equivalent smooth case value and its corresponding near wall peak $u'_{g,w_p}$, which actually is lower than in the smooth case. Both observations are direct consequences of the presence of waves. The enhancement at the interface is easily explained by the fact that, in a time-averaged sense, the gas phase perceives the wavy interface as a moving wall with a considerable roughness. Higher roughness leads inevitably to thicker and more turbulent boundary-layer near the interface.

However, the latter observation (drop near the wall) may seem rather surprising and its explanation is a bit more complicated. Berthelsen and Ytrehus (2005), have actually documented this phenomenon indirectly through turbulent kinetic energy considerations. They developed a numerical method for calculation of wavy two-phase flow where the wavy interface is represented by an equivalent interfacial roughness, in accordance with the 'moving rough wall'- argument made above. It is reasonable to make an analogy between the axial turbulence level and the turbulent energy level inside a boundary layer, since most of the energy inside such a layer is concentrated in the axial direction. This can be seen by comparing the peak values of the measured $u'$ and $v'$-profiles in the present study, and by assuming that the azimuthal component $w'$ is comparable to $v'$, as seen in van Doorne and Westerweel's work (2007). 

A simple dimensional analysis shows that the near-wall axial turbulent energy peak exceeds its equivalent wall-normal and azimuthal components with a factor 25. According to the DNS study of Lakehal et. al (2003), the interface deformations onset a redistribution of energy by reducing the dissipation terms while leaving the production terms unchanged. This implies that the turbulent activity persists near the interface, which is confirmed by this study.

An alternative explanation of the physics that lie behind the reduction of the wall axial turbulence may be obtained by considering the following thought experiment: It is of common knowledge that the onset of waves in a gas-liquid flow is caused by an instability or a perturbation at the interface that grows with time until it reaches a stable wavy state 
which is maintained by the kinetic energy of the gas phase. Theoretically, one could prevent such instabilities from occurring such that the interface would remain smooth even at gas flow rates that normally would generate a waves.This could be achieved in a perfectly round, smooth and horizontal pipe which is isolated from all temperature and pressure changes and in which both the gas and liquid flows are perfectly isotropic and homogeneous. Obviously, these conditions are rather impossible to attain in a laboratory. 

Another important feature of stratified gas-liquid pipe flow that must be recalled, is that the turbulent energy production is mainly concentrated inside the viscous boundary layers near the wall and the interface. In the special case of a smooth interface flow, it can be argued that production rate is higher inside the wall boundary layer than in the interfacial boundary layer since the stream wise velocity gradient is lower there. 

Now, by introducing a perturbation to the interface, the flow would quickly stabilize into a wavy regime with characteristics conformed by the kinetic energy that already exists in the gas phase. Considering this case in a time-averaged sense, the 'moving rough wall'- analogy of the interface would imply higher turbulent energy production rate in the interfacial boundary. Since no external energy was added to the system, a redistribution of energy must take place in the remaining pipe region, including the wall region. This is in agreement with both the drop seen in figure \ref{Fig::PIV_2Phase_2Runs_Profiles_May2012} and Lakehal's conclusion (2003).

Furthermore, a dramatic enhancement of the liquid wall-normal turbulent component $v'_l$ is seen in the wavy case. Qualitatively, this observation is not surprising. However, extra care is needed when evaluating this result quantitatively. It should be remarked that the oscillatory motion of the waves must have had an influence on the computation of the $v$-rms profile. One wayto deal with this challenge, is to subtract the oscillations due to the waves from the raw wall-normal velocity profile before computing the rms.

These qualitative results confirm many already known aspects of the turbulence structure in stratified gas/liquid two-phase flow and offer a motivation for further investigation.

\begin{figure}[h!]
\begin{center}
\includegraphics[width=\columnwidth, height=0.5\textheight]{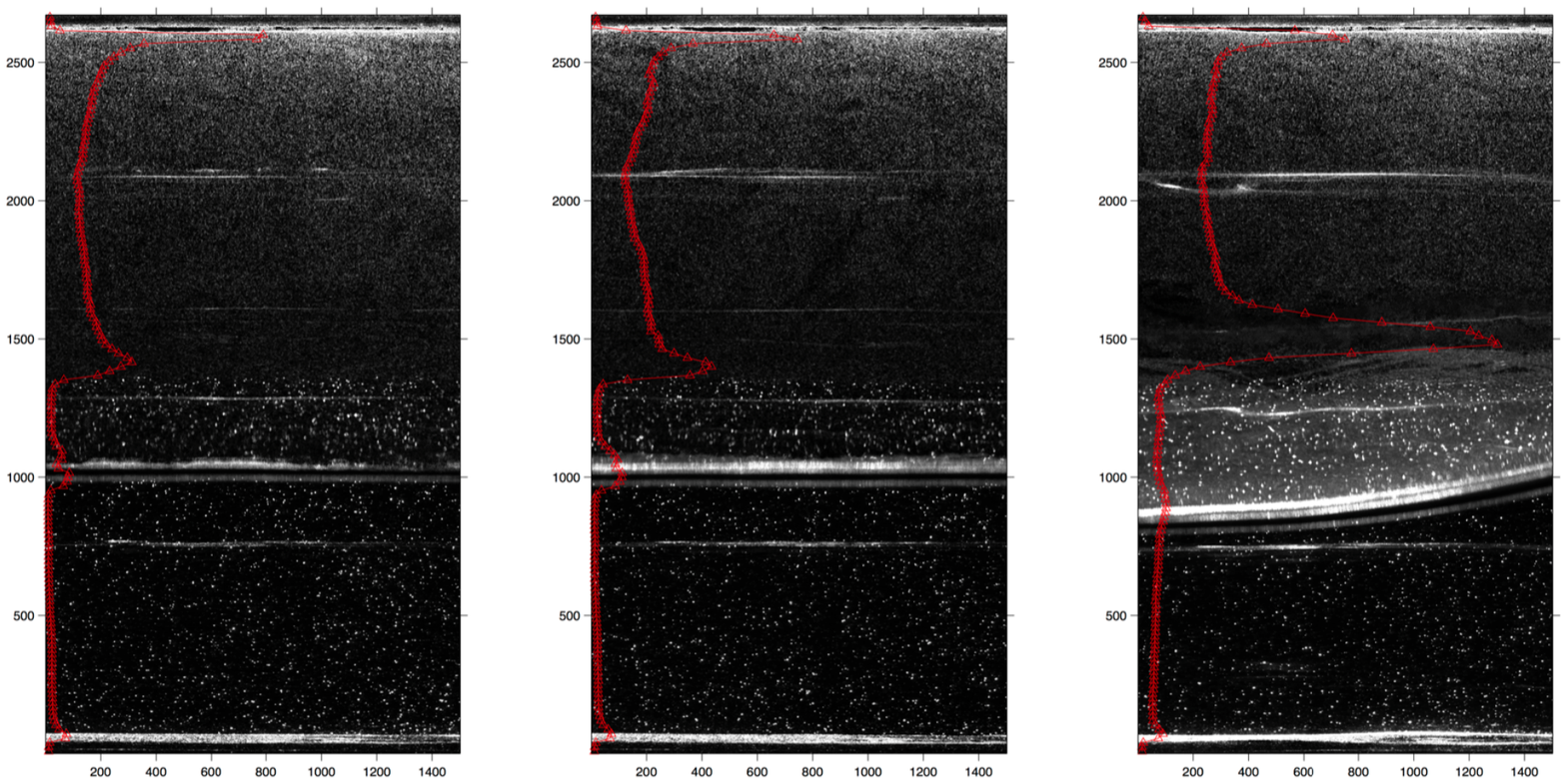}\\
\caption{\small PIV images of a smooth, rippled and wavy flow shown with their belonging (time-averaged) axial turbulence profiles.} \label{Fig::3Images}
\end{center}
\end{figure}

\subsection{Streamwise Turbulence Analysis}
\label{sec::STA}

A set of 17 stratified air-water flow experiments of varying flow rates were conducted using the PIV set-up described earlier. Measured physical and flow properties such as the liquid hold-up $H_L$, the gas pressure drop $\frac{\Delta P_g}{L}$, the gas and liquid Re-numbers (both superficial and hydraulic), the mean and superficial velocities, $U_{b,f}$ and $U_{sf}$ and finally, the axial gas turbulence profiles are investigated further. The focus of this study is limited to the axial turbulence, partially because a great deal of consistency was found in all $u'$-profiles, confirming the high degree of accuracy in second order statistics in this direction, and also because the shape of the profiles looked to cohere with the nature of the interface, as seen in figure \ref{Fig::3Images}, hinting about the existence of a correlation between them. Hence, the aim of this study is to search for such a correlation, at least qualitatively. The table in figure \ref{Tab::Re-numbers} contains the physical parameters of each experiment, 
while figure \ref{Fig::TurbAnalysis_ReDg_vs_HL} displays their liquid hold-ups. 

Different flow patterns were identified visually while performing the experiments and these were found to match the descriptions provided by Espedal (1998), Strand (1993) and 
Fernandino and Ytrehus (2006). Espedal and Strand used wave probes inside their pipes, respectively 100 mm and 60 mm i.d., and analyzed the cross correlation values in order to quantify the various wave patterns. Espedal divided them into the following 5 different regions:

\ben
\item Smooth flow: No waves were observed.
\item Small amplitude waves I: Amplitudes below 2 mm and wave lengths between 2 and 6 cm. The power spectrum showed no peak at all or one peak.
\item Small amplitude 2D waves II: Similar to the waves above, but the power spectrum showed two peaks.
\item Large amplitude 2D waves: Amplitudes above 2 mm, and the waves are less regular. The power spectrum has a one, two or no marked peaks.
\item Large amplitude 3D waves: Amplitudes above 2 mm, and the waves do not have a two dimensional shape.
\een

Ytrehus and Fernandino (2006) used the laser doppler velocimetry (LDV) technique to measure the fluctuations that occur close to the air-sheared interface. By analyzing the spectra of the LDV signal, they emphasized that intermediate patterns exist between the well defined ones. Espedal's identifications up to the "small amplitude 2D waves II" were found to agree very well with the flow patterns that were observed during this study. However, the second sub-regime is referred to as "rippled flow" and is believed to consist of capillary waves riding on top of an otherwise flat interface. The rippled flow shares some common features with the smooth flow in regard to the turbulence structure. This is discussed in more details later. 

\begin{figure}[ht!]
\begin{center} 
\begin{minipage}[t]{0.48\columnwidth}
\centering
\includegraphics[width=\columnwidth, height=0.3\textheight]{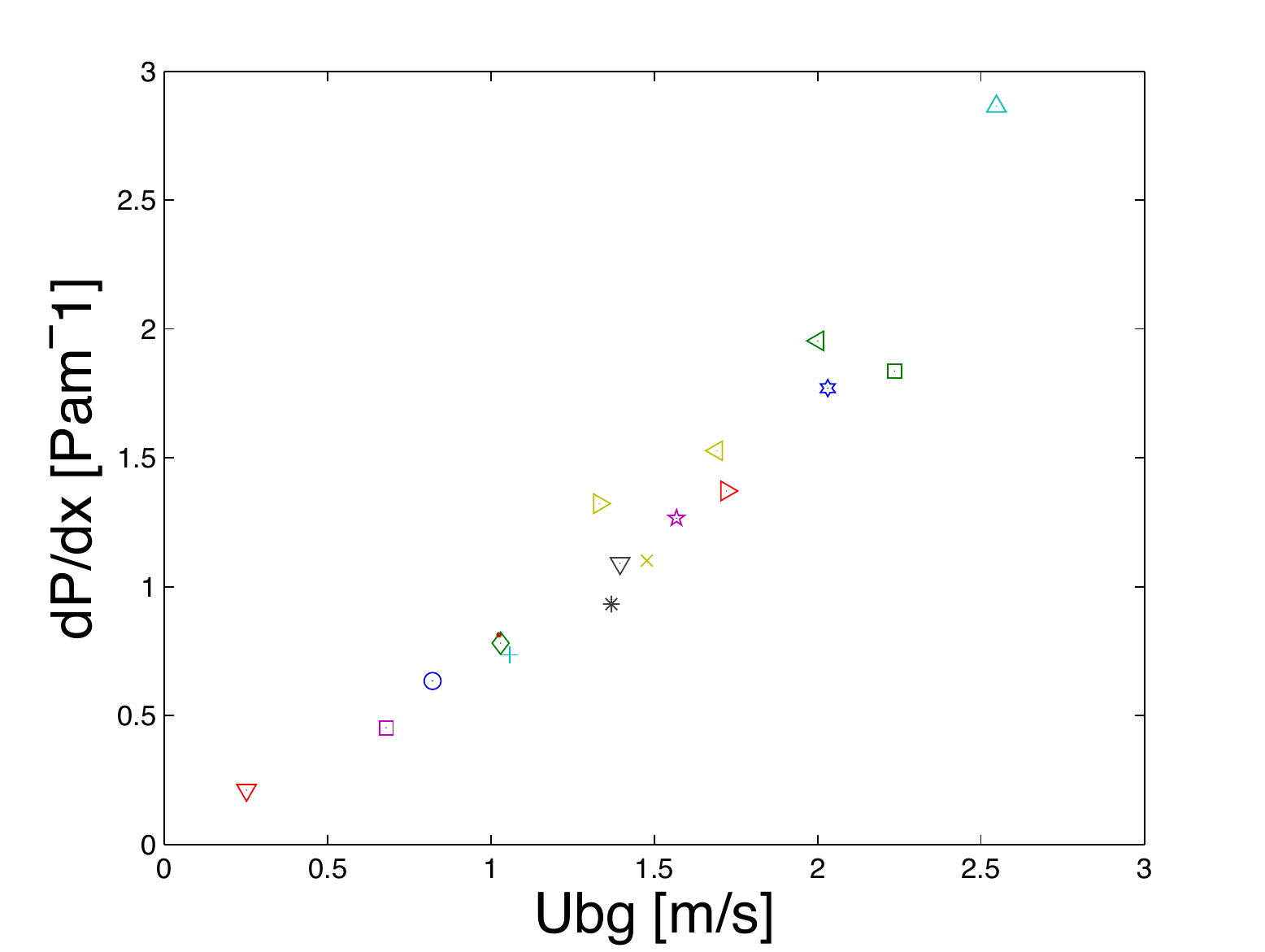}
\caption{\small The measured air pressure drop $\frac{\Delta P}{L}$ plotted against $Ub_g$} \label{Fig::TurbAnalysis_Ubg_dPdx}
\end{minipage}
\hspace{0.15cm}
\begin{minipage}[t]{0.48\columnwidth}
\centering
\includegraphics[width=\columnwidth, height=0.3\textheight]{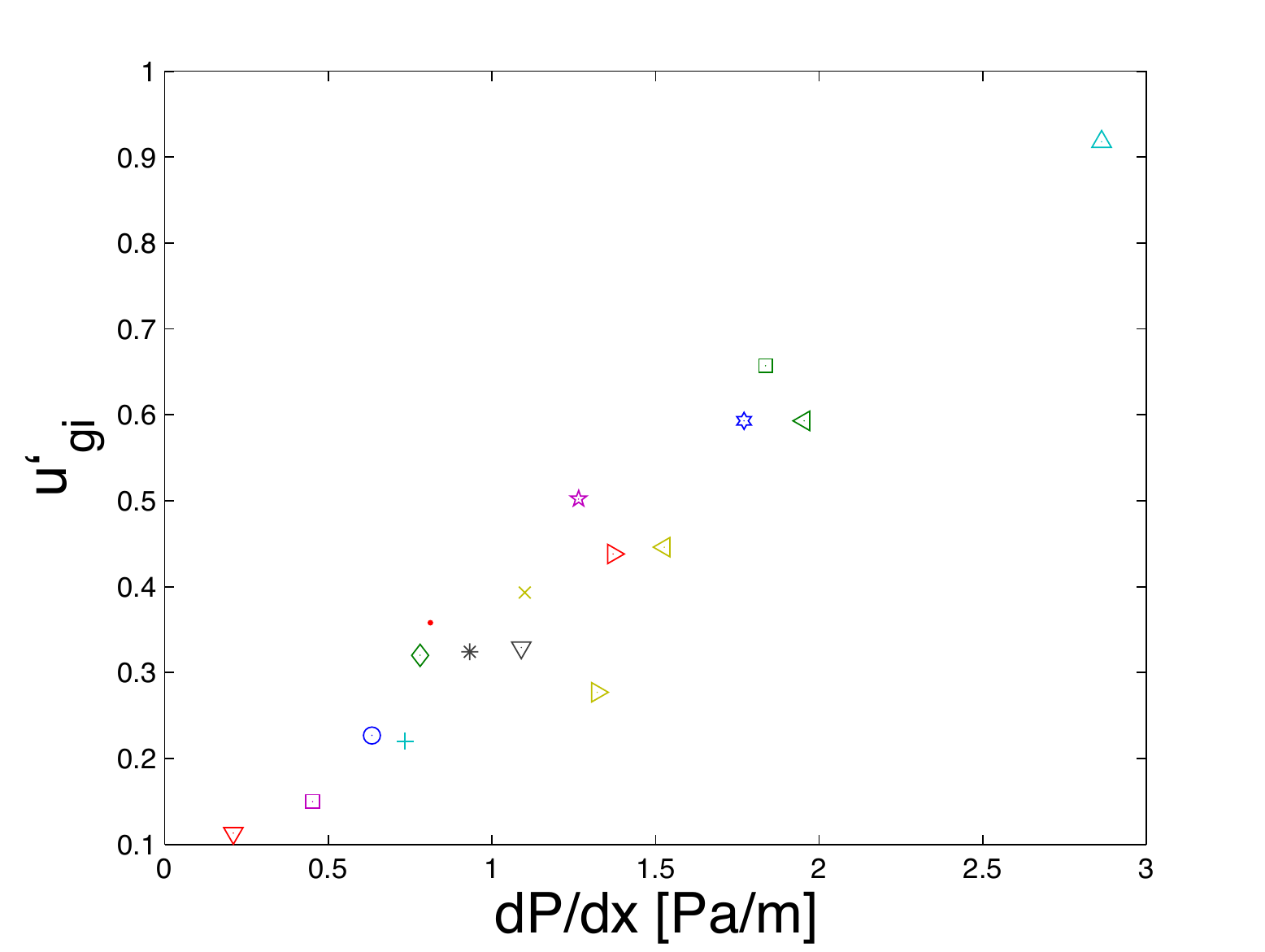}
\caption{\small The axial turbulence peak near the interface $u'_{g,i_p}$ plotted against $dP/dx$.} \label{Fig::TurbAnalysis_ugi_dPdx}
\end{minipage}
\end{center}
\end{figure}

Figure \ref{Fig::TurbAnalysis_Ubg_dPdx} shows the measured gas pressure drops of each experiment plotted with their corresponding mean local velocities $U_{bg}$. There seems to be a linear relationship between them, at least for the relatively low pressure region in which the present experiments operate. Strand (1993) claimed that the pressure drop increases almost quadratically with respect to the gas flow rate for fixed liquid rates. But, this does not have to contradict the present results, since he studied flows with significantly higher pressure drops.  

\begin{figure}[ht!]
\begin{center} 
\begin{minipage}[t]{0.48\columnwidth}
\centering
\includegraphics[width=\columnwidth, height=0.3\textheight]{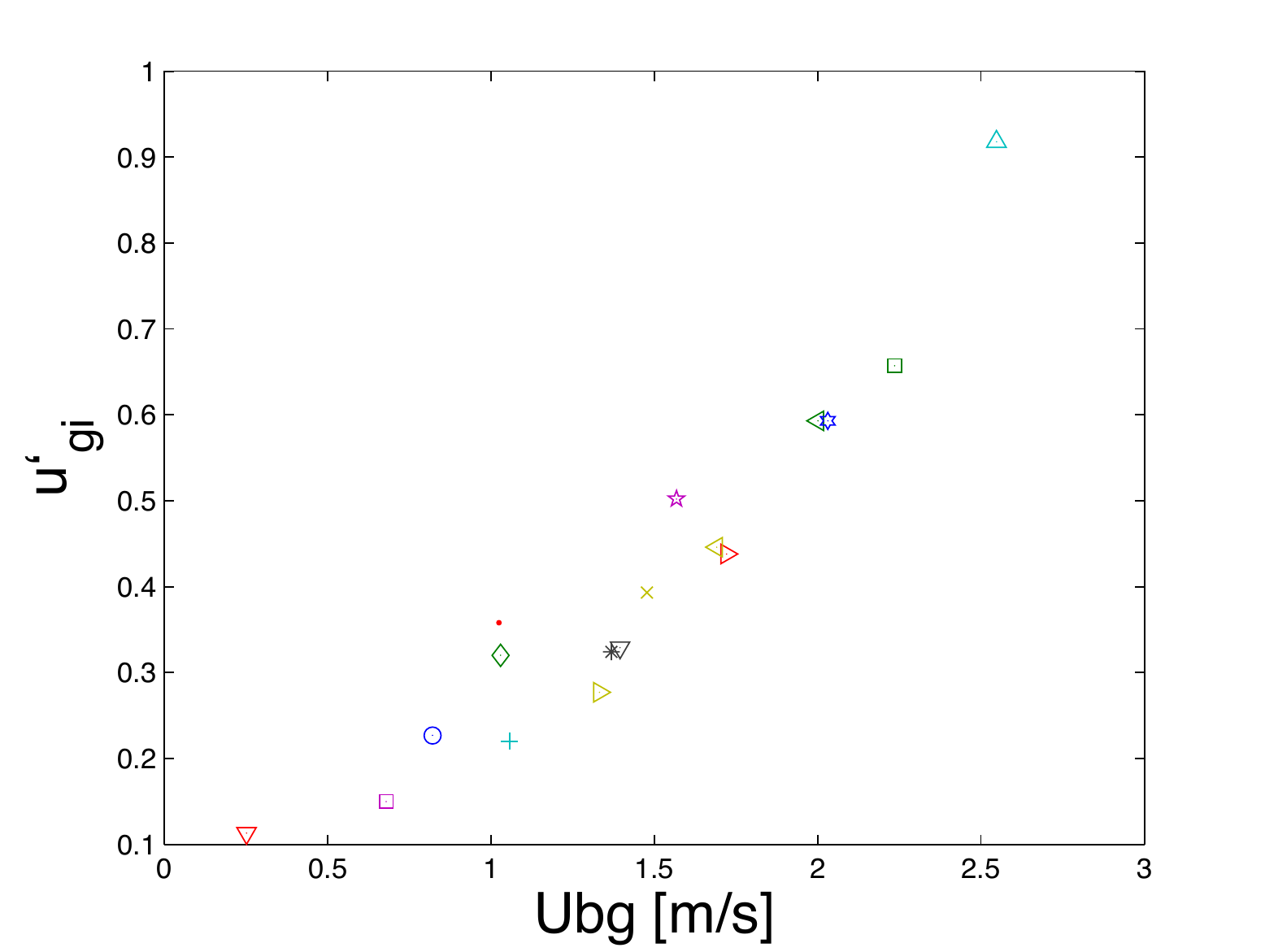}
\caption{\small The axial turbulence peak near the interface $u'_{g,i_p}$ plotted against $Ub_g$} \label{Fig::TurbAnalysis_Ubg_vs_ugi}
\end{minipage}
\hspace{0.15cm}
\begin{minipage}[t]{0.48\columnwidth}
\centering
\includegraphics[width=\columnwidth, height=0.3\textheight]{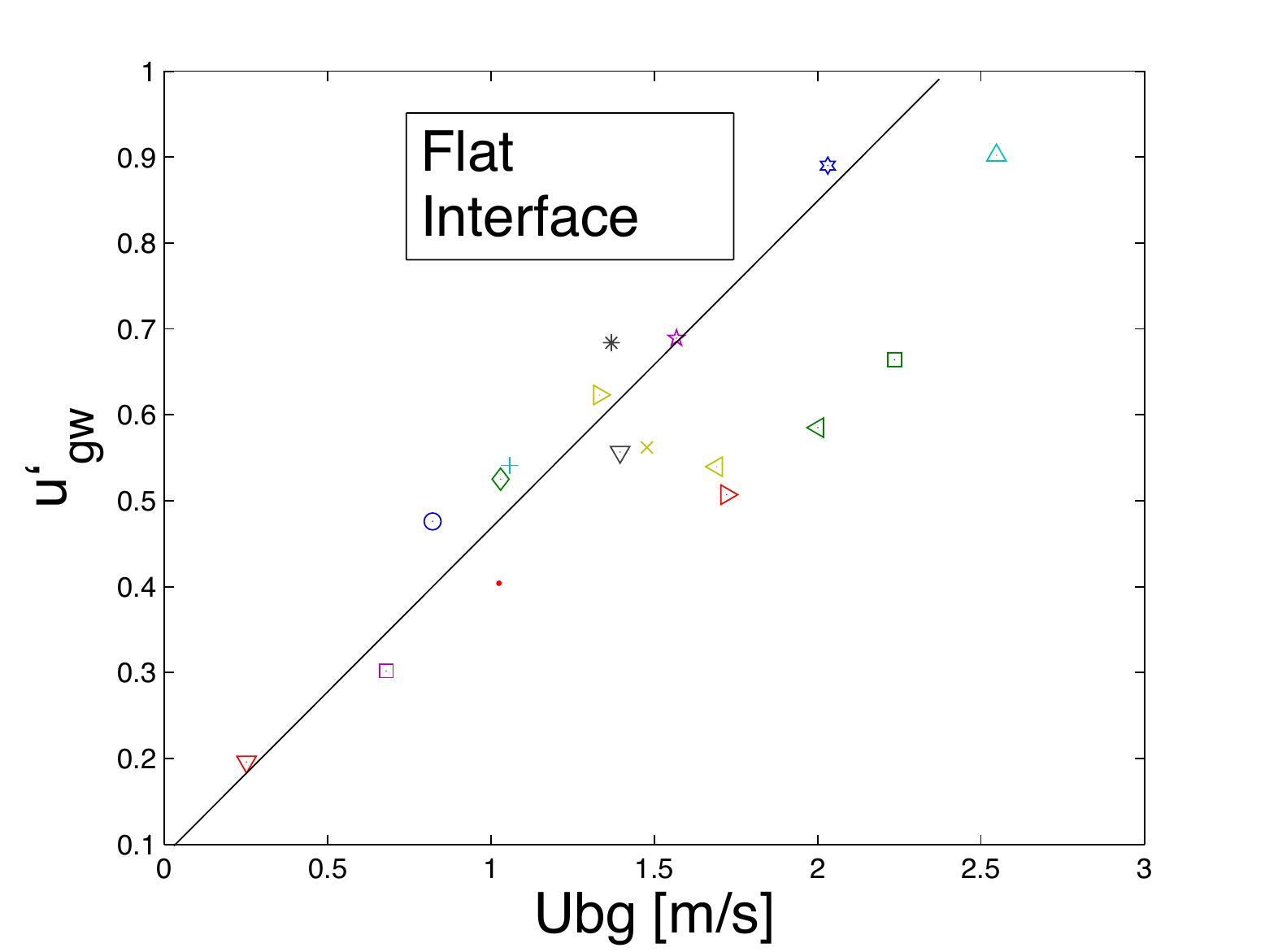}
\caption{\small The axial turbulence peak near the wall $u'_{g,w_p}$ plotted against $Ub_g$} \label{Fig::TurbAnalysis_Ubg_vs_ugw}
\end{minipage}
\end{center}
\end{figure}

One of the most important results of this study is seen in figure \ref{Fig::TurbAnalysis_ugi_dPdx}, where the near interface peak value of the axial turbulence, $u_{g,i_p}$, is plotted against the pressure drop. Here, a very similar linear correlation is found as in figure \ref{Fig::TurbAnalysis_Ubg_dPdx}. This result indicates that the same relationship should be expected also between $u_{g,i_p}$ and $U_{bg}$. This is seen in figure \ref{Fig::TurbAnalysis_Ubg_vs_ugi}. This means that changes in the local mean velocity affect the pressure drop and the interfacial axial turbulence analogously. 

Furthermore, figure \ref{Fig::TurbAnalysis_Ubg_vs_ugw} shows that in the smooth and rippled interface cases (generalized by the term "flat"), the wall turbulence also follows a linear evolution with respect to increasing $U_{bg}$. However, a drop from this linear trend is present in all non-flat cases, confirming the qualitative observations made in the previous section. This result explains the choice of the term "rippled" instead of small amplitude 2D waves as denoted by Espedal. The capillary waves have very small amplitudes, and the interface deformations that they cause are not sufficiently large for a redistribution of energy to occur.

\begin{figure}[ht!]
\centering
\includegraphics[width=0.8\columnwidth, height=0.4\textheight]{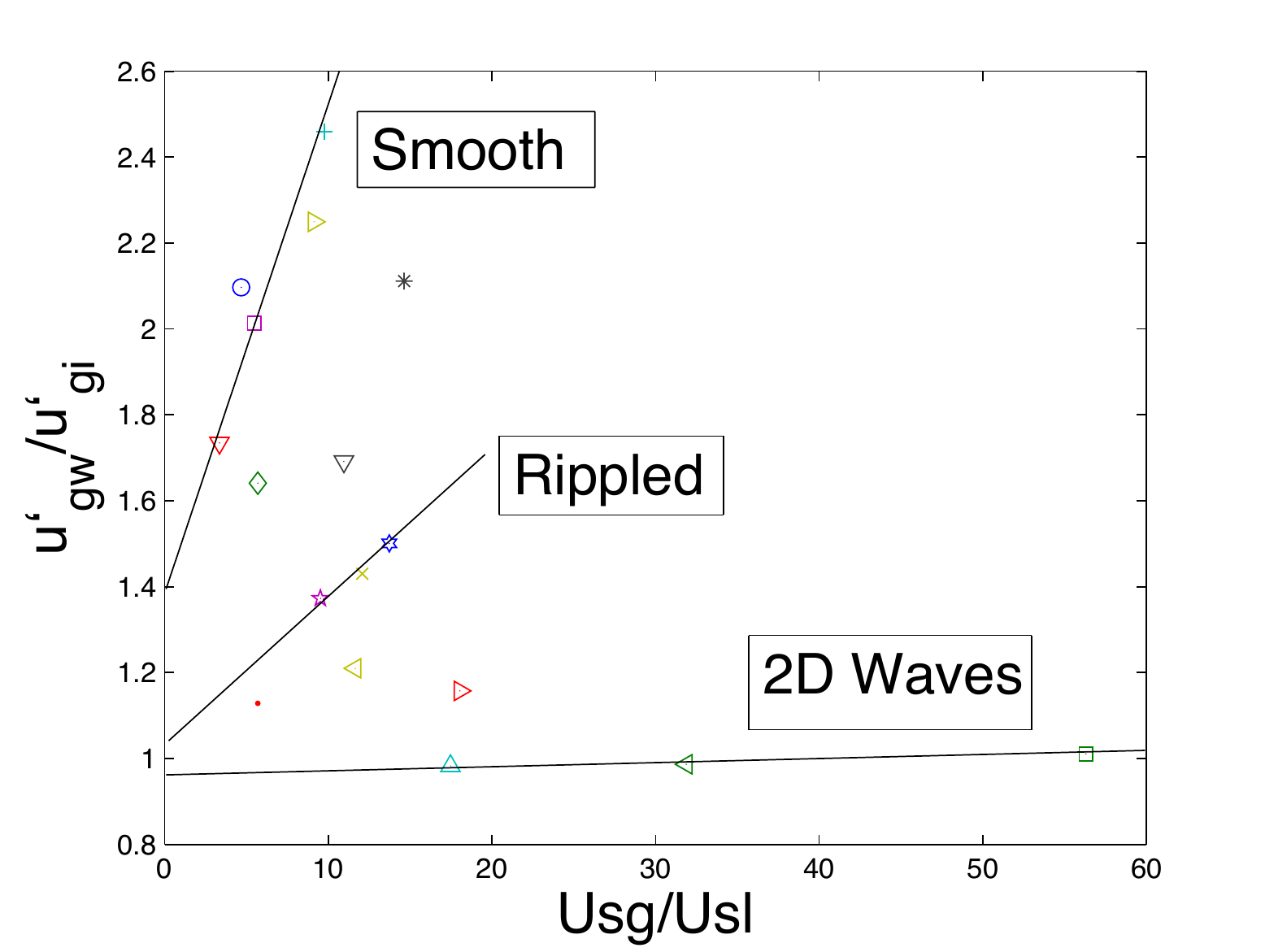}
\caption{\small $u'_{g,w_p}/u'_{g,i_p}$ plotted against $U_{sg}/U_{sl}$} \label{Fig::TurbAnalysis_Usg_Usl_vs_ugw_ugi}
\end{figure}

Finally, the main result of this paper is presented in figure \ref{Fig::TurbAnalysis_Usg_Usl_vs_ugw_ugi}. Here the ratio between the wall and interface turbulence peaks $\frac{u'_{g,w_p}}{u'_g,i_p}$ is plotted against the superficial velocity ratio $\frac{U_{sg}}{U_{sl}}$. A clear distribution of three linear trends arise in accordance to the observed flow patterns, smooth, rippled and small amplitude 2D waves. Some intermediate cases are also visible between the linear trends. This is in agreement with Fernandino's (2006) conclusions. This result proves that the turbulence structure of any stratified gas-liquid flow is characterized by its actual interface behavior.

\section{Concluding remarks}
\label{sec::conclusion}
Simultaneous PIV measurements of stratified gas/liquid two-phase flow in a horizontal pipe have been successfully performed using water droplets, $1-5 \mu m$ in diameter as tracers in the gas phase. Single phase experiments have been conducted for the sake of validation of the seeding technique and the experimental set-up. The results from these have been compared to DNS results and to liquid single phase experiments showing perfect agreement in the first order statistics, i.e. $\bar U$-profile, and satisfactory agreement in the second order statistics, i.e. $u'$, $v'$ and $u'v'$ profiles. 

The two-phase flow measurements showed expected behavior for low flow rates, and very interesting features related to the turbulence structure at higher flow rates that led to wavy interfaces. These can be summarized as follows:

\bit
\item A smooth interface is perceived as a moving wall by the gas-phase. Hence a higher axial turbulence level is found near the wall than near the interface.
\item A wavy interface causes an enhancement of the turbulence level close to the interface. This is because, in a time-averaged sense, the waves act as a rough wall.
\item The axial turbulence level near the wall drops in the transition to wavy flow due to a redistribution of the turbulent kinetic energy. 
\eit

A relatively extensive study consisting of 17 different stratified two-phase flow experiments has shown that the pressure drop, the mean velocity and the interfacial axial turbulence level all are linearly coupled. At least at the low flow rates in which the present experiments operate. The main result of this work showed that there exists a qualitative correlation between the turbulence structure of a given stratified two-phase flow and its corresponding interface flow pattern. This is an interesting that result may eventually provide valuable tools in two-phase flow modeling, where some models link the turbulence structure to the friction factors, e.g. Biberg (2007).

\end{document}